\documentclass[prd,aps,showpacs,epsf,floats,10pt]{revtex4}
\usepackage{amssymb}
\usepackage{amsfonts}
\usepackage{amsmath}

\input{tcilatex}
\begin{document}

\title{ON A DIFFERENTIAL GEOMETRIC VIEWPOINT OF JAYNES' MAXENT METHOD AND
ITS QUANTUM EXTENSION}
\author{S. A. Ali$^{1,2}$, Carlo Cafaro$^{3}$, Adom Giffin$^{4}$, Cosmo Lupo$%
^{3}$, Stefano Mancini$^{3}$}
\affiliation{$^{1}$International Institute for Theoretical Physics and Mathematics
Einstein-Galilei, via Santa Gonda 14, 59100 Prato, Italy\\
$^{2}$Department of Arts and Sciences, Albany College of Pharmacy and Health
Sciences, 106 New Scotland Avenue, Albany, NY 12208, USA\\
$^{3}$School of Science and Technology, Physics Division, University of
Camerino, I-62032 Camerino, Italy\\
$^{4}$Princeton Institute for the Science and Technology of Materials,
Princeton University, Princeton, NJ 08540, USA}

\begin{abstract}
We present a differential geometric viewpoint of the quantum MaxEnt estimate
of a density operator when only incomplete knowledge encoded in the
expectation values of a set of quantum observables is available. Finally,
the additional possibility of considering some prior bias towards a certain
density operator (the prior) is taken into account and the unsolved issues
with its quantum relative entropic inference criterion are pointed out.
\end{abstract}

\maketitle

\section{Introduction}

It is well established that \emph{geometry} plays an important role in
characterizing and understanding both classical and quantum physics. After
all, it has been an old dream to reduce the fundamental laws of physics to
geometry since Einstein's formulation of general relativity. In particular,
it is a remarkable achievement that all the building blocks of quantum field
theory can be formulated in terms of geometric concepts such as vector
bundles, connections, curvatures, covariant derivatives and spinors \cite%
{frankel}.

It is well known that all practical problems in science are characterized by
incompletely specified situations where we can only provide an \emph{%
inference} (plausible conjecture) \cite{jaynes1}: given incomplete
information about reality, we\textbf{\ }choose probability distributions
that represent the state of (incomplete) knowledge of the system being
considered. The incompleteness of information leads to the non-uniqueness of%
\textbf{\ }the reconstructibility of the state of the classical system, that
is, more than one situation is compatible with what is known. However, some
situations are more likely to occur than others. This leads to a
probabilistic description of dynamical systems in the presence of partial
knowledge. Furthermore, it is also the case that geometry can be of some use
for the study of inference. In particular, a number of investigations have
shown that a useful approach to the study of statistical inference is to
regard a parametric statistical model as a differentiable manifold equipped
with a metric structure \cite{brody}. Of course, as pointed out by Skilling 
\cite{skilling}, like any other professional tool, "\emph{geometry should be
used with intelligence and care}".

In recent years, we have employed information geometry \cite{amari}
(Riemannian geometry applied to probability theory) and inductive inference
(the Maximum relative Entropy method (MrE) \cite{caticha-giffin} of which
Jaynes' MaxEnt \cite{jay} is a special case) to study the complexity of
informational geodesic flows on curved statistical manifolds (statistical
models) underlying the probabilistic description of physical systems in the
presence of incomplete information \cite{carlo}. Most of this work has dealt
with ordinary probability distributions. However, what happens if the state
of the system is represented by a quantum density operator? Are Riemannian
geometric techniques of some utility in such cases? Is there an inference
method useful to assign a density operator given a specific set of
expectation values of quantum observables? Is there a quantum entropic
inference method that can be used to update density operators when some
prior bias towards a certain density operator (the prior) is assumed? Such
inference problems occur in situations in which information is gained in two
steps. In the first step, experiments are carried out to determine a prior
density operator $\rho _{0}$. In the second step, more knowledge is gained
about the expectation values of certain quantum observables $\left\{
A_{k}\right\} $. How should one determine the density operator $\rho $ which
is best suited to make further predictions, accounting fully for both types
of information? In this work we provide a few answers to such questions and
emphasize some conceptual and computational problems that may emerge.

\section{Inference and Density Operators}

Jaynes's MaxEnt method can be extended to the quantum density-matrix
formalism in a straightforward manner \cite{jaynes2}.

Assume we are given the expectation values of the quantum operators $A_{1}$%
,..., $A_{m}$. Stated otherwise, the quantities $A_{j}$ with $j=1$,..., $m$
are the quantum observables which represent the\textbf{\ }slow variables of
the theory \cite{grasselli}, that is, those whose expectation values can be
measured at any given time. For the sake of reasoning, assume that the
information constraints are given by,%
\begin{equation}
\bar{A}_{j}=\left\langle A_{j}\right\rangle \overset{\text{def}}{=}\text{tr}%
\left( \rho A_{j}\right) \text{, with }j=1\text{,..., }m\text{.}  \label{ev}
\end{equation}%
We point out that if the constraints (\ref{ev}) involve only commuting
observables $A_{j}$, the quantum method can be reduced to the classical one.
In this case, there exists a basis in the Hilbert space in which all $A_{j}$
are represented by diagonal matrices. In this basis, $\rho $ should also be
diagonal and diagonal elements are ordinary probabilities. Indeed, it can be
shown that Jaynes' MaxEnt method also works for non-commuting observables 
\cite{bala}. In this case, however, the off-diagonal elements of $\rho $
must be considered and this has no equivalent in standard probability
theory. The values $\bar{A}_{j}$ are averages obtained in a large number of
experiments performed on an ensemble of identically prepared systems. The
necessity of an ensemble is obvious if we consider a situation in which the
state to be characterized is specified by giving the expectation values of
non-commuting observables. This requires that measurements associated with
each (non-commuting) observable is performed on different samples, belonging
to separate sub-ensembles of the full ensemble whose state is described by $%
\rho $. Using the Lagrange multipliers technique, it can be shown that the
density matrix $\rho $ which represents the most unbiased picture of the
state of the system on the basis of this much information is the one
maximizing the von Neumann entropy $\mathcal{S}(\rho )$,%
\begin{equation}
\mathcal{S}(\rho )\overset{\text{def}}{=}-\text{tr}(\rho \log \rho )\text{,}
\label{vn}
\end{equation}%
subject to the information constraints in (\ref{ev}) and to the
normalization condition tr$(\rho )=1$ . As a side remark, we point out that $%
\mathcal{S}(\rho )$ is invariant under unitary transformations. This is an
important property of the von Neumann entropy and is very desirable since we
require that our macroscopic predictions should remain unchanged if the
observables are unitarily transformed. As in the classical case, the
maximization process is accomplished by finding the density matrix $\hat{\rho%
}$ which solves the variational problem,%
\begin{equation}
\delta \left[ -\text{tr}\left( \rho \log \rho \right) -\lambda _{0}\left( 
\text{tr}\rho -1\right) -\sum_{k=1}^{m}\lambda _{k}\left( \text{tr}\left(
\rho A_{k}\right) -\bar{A}_{k}\right) \right] =0\text{,}  \label{vp}
\end{equation}%
where $\delta =\delta _{\rho }$ is the variational operator. After some
simple algebra, it follows that Eq. (\ref{vp}) reads,%
\begin{equation}
\left( \delta \rho \right) \left[ \log \rho +\lambda _{0}\mathbf{1}%
+\sum_{k=1}^{m}\lambda _{k}A_{k}\right] +\rho \left( \delta \left[ \log \rho
+\lambda _{0}\mathbf{1}+\sum_{k=1}^{m}\lambda _{k}A_{k}\right] \right) =0%
\text{,}  \label{vip1}
\end{equation}%
where $\mathbf{1}$ is the identity operator. We point out that one of the
main technical difficulties that occur in the quantum setting is that the
chain rule for derivatives does not hold: $\rho $ and $\delta \rho $ do not
necessarily commute \cite{grasselli}. That said, the solution of the
variational problem in (\ref{vip1}) is achieved by imposing the condition,%
\begin{equation}
\log \rho +\lambda _{0}\mathbf{1}+\sum_{k=1}^{m}\lambda _{k}A_{k}=0\text{,}
\end{equation}%
i.e.,%
\begin{equation}
\rho =\exp \left[ -\lambda _{0}\mathbf{1}-\sum_{k=1}^{m}\lambda _{k}A_{k}%
\right] \text{.}  \label{ro}
\end{equation}%
The Lagrange multipliers $\lambda _{0}$ in (\ref{ro}) is determined by the
normalization condition tr$\left( \hat{\rho}\right) =1$. As a matter of
fact, noticing that $\exp \left( -\lambda _{0}\mathbf{1}\right) $ equals $%
\exp \left( -\lambda _{0}\right) \mathbf{1}$, the condition tr$\left( \rho
\right) =1$ yields%
\begin{equation}
1=\text{tr}\left[ \exp \left( -\lambda _{0}\mathbf{1}-\sum_{k=1}^{m}\lambda
_{k}A_{k}\right) \right] =\exp \left( -\lambda _{0}\right) \text{tr}\left[
\exp \left( -\sum_{k=1}^{m}\lambda _{k}A_{k}\right) \right] \text{,}
\end{equation}%
that is,%
\begin{equation}
\lambda _{0}=\log \mathcal{Z}\left( \lambda _{1}\text{,..., }\lambda
_{m}\right) \text{,}  \label{lambda0}
\end{equation}%
where $\mathcal{Z}$ is the so-called partition function defined as,%
\begin{equation}
\mathcal{Z}=\mathcal{Z}\left( \lambda _{1}\text{,..., }\lambda _{m}\right) 
\overset{\text{def}}{=}\text{tr}\left[ \exp \left( -\sum_{k=1}^{m}\lambda
_{k}A_{k}\right) \right] \text{.}
\end{equation}%
The remaining Lagrange multipliers $\lambda _{k}$ with $k=1$,..., $m$ are
determined by the information constraints (\ref{ev}). In particular one
finds,%
\begin{equation}
\bar{A}_{j}=\left\langle A_{j}\right\rangle \overset{\text{def}}{=}\text{tr}%
\left( \rho A_{j}\right) =\exp \left( -\lambda _{0}\right) \text{tr}\left[
A_{j}\exp \left( -\sum_{k=1}^{m}\lambda _{k}A_{k}\right) \right] =\frac{1}{%
\mathcal{Z}\left( \lambda _{1}\text{,..., }\lambda _{m}\right) }\left( -%
\frac{\partial \mathcal{Z}\left( \lambda _{1}\text{,..., }\lambda
_{m}\right) }{\partial \lambda _{j}}\right) \text{,}
\end{equation}%
that is,%
\begin{equation}
\bar{A}_{j}=-\frac{\partial \log \mathcal{Z}\left( \lambda _{1}\text{,..., }%
\lambda _{m}\right) }{\partial \lambda _{j}}\text{.}  \label{rel1}
\end{equation}%
Thus, the Lagrange multipliers $\lambda _{j}$ with $j=1$,..., $m$ are
implicitly defined in Eq. (\ref{rel1}). Indeed, the multipliers $\lambda
_{j} $ may also be expressed in an explicit manner. Observing that $S_{\max
}=S\left( \rho \right) $ with $\rho $ in (\ref{ro}) is given by,%
\begin{equation}
\mathcal{S}_{\max }=\mathcal{S}_{\max }\left( \bar{A}_{1}\text{,..., }\bar{A}%
_{m}\right) =\lambda _{0}+\lambda _{1}\bar{A}_{1}+\text{....}+\lambda _{m}%
\bar{A}_{m}\text{,}
\end{equation}%
we conclude that,%
\begin{equation}
\lambda _{j}=\frac{\partial \mathcal{S}_{\max }\left( \bar{A}_{1}\text{,..., 
}\bar{A}_{m}\right) }{\partial \bar{A}_{j}}\text{.}  \label{lambdaj}
\end{equation}%
In general, it may be highly non trivial to solve explicitly the system of
equations for the Lagrange multipliers $\lambda _{j}$\ in (\ref{lambdaj}).
Finally, substituting (\ref{lambda0}) and (\ref{lambdaj}) into (\ref{ro}),
we obtain the "\emph{quantum MaxEnt estimate}",%
\begin{equation}
\rho =\rho \left( A_{1}\text{,..., }A_{m}|\bar{A}_{1}\text{,..., }\bar{A}%
_{m}\right) =\frac{\exp \left[ -\overset{m}{\underset{k=1}{\sum }}\frac{%
\partial S_{\max }\left( \bar{A}_{1}\text{,..., }\bar{A}_{m}\right) }{%
\partial \bar{A}_{k}}A_{k}\right] }{\text{tr}\left[ \exp \left( -\overset{m}{%
\underset{k=1}{\sum }}\frac{\partial S_{\max }\left( \bar{A}_{1}\text{,..., }%
\bar{A}_{m}\right) }{\partial \bar{A}_{k}}A_{k}\right) \right] }\text{.}
\label{estimate}
\end{equation}%
We point out that the quantum MaxEnt estimate is always a physical state
thanks to the canonical form of the density operator in (\ref{estimate}). We
recall that although Jaynes' MaxEnt was originally developed for assigning
probability distributions \cite{jay}, it can also be regarded as a special
case of the MrE method when updating probability distributions from a
uniform prior using the Gibbs-Shannon entropy \cite{caticha-giffin}. In the
quantum framework, Jaynes' MaxEnt is employed to assign density operators
via maximization of the von Neumann entropy \cite{jaynes2}. However,
although at the moment there is no universal quantum MrE method for updating
density operators, the quantum MaxEnt formalism may be viewed as a limiting
case of a possible updating criterion from uniform prior density operators
where inference is carried out by means of the logarithmic quantum relative
entropy,%
\begin{equation}
\mathcal{S}\left( \rho ||\rho _{0}\right) \overset{\text{def}}{=}-\text{tr}%
\left[ \rho \left( \log \rho -\log \rho _{0}\right) \right] \text{.}
\label{lqre}
\end{equation}%
When the prior $\rho _{0}$\ is uniform and proportional to the identity
operator $1$\ acting on an\ $n$-dimensional Hilbert space, $S\left( \rho
||\rho _{0}\right) $ reads%
\begin{equation}
\mathcal{S}\left( \rho ||\rho _{0}\right) =\mathcal{S}\left( \rho \right)
-\log n\text{.}  \label{vng}
\end{equation}%
Thus, maximizing (\ref{vn}) is equivalent to maximizing (\ref{vng}). We
remark that the use of (\ref{lqre}) as a suitable entropic tool for quantum
inferences (updating density matrices from arbitrary priors) appears in \cite%
{private} as well. For forthcoming utility, we assume a bias towards a
uniform prior density operator $\rho _{0}$. Then, the maximization of the
quantum logarithmic relative entropy (\ref{vng}) subject to a fixed
information constraint $\bar{A}=\left\langle A\right\rangle $\ leads to the
MaxEnt estimate,%
\begin{equation}
\rho \left( \lambda \right) =\rho _{0}\frac{e^{-\lambda A}}{\text{tr}\left(
e^{-\lambda A}\rho _{0}\right) }=\frac{e^{-\frac{\lambda }{2}A}\rho _{0}e^{-%
\frac{\lambda }{2}A}}{\text{tr}\left( e^{-\lambda A}\rho _{0}\right) }\text{.%
}  \label{maxent-estimate}
\end{equation}%
The Jaynes MaxEnt method has been successfully applied to partial
(incomplete) reconstruction of density operators of quantum systems from the
available measured mean values of the system's observables (single spin-$%
\frac{1}{2}$, two correlated spins-$\frac{1}{2}$,
Greenberger-Horne-Zeilinger states, vibrational states of neutral atoms,
etc.) \cite{buzek}. Unfortunately, Jaynes' MaxEnt quantum formalism is not
without criticisms in the scientific community. Indeed, it has been claimed
that quantum state estimates derived via the principle of MaxEnt are
fundamentally different from those obtained via the quantum Bayes rule \cite%
{schack}. This seems understandable and it is somehow expected. By
definition, "a probability is an abstract notion that represents the degree
of plausibility of a proposition, subject to information regarding that
proposition" \cite{giffin}. Thus if one has different information one should
come to different probabilities. However, just as Maximum relative Entropy
successfully showed that one can use a universal method (MrE) for any type
of classical information, it may be possible that a quantum version of MrE
will do the same for the quantum case.

\section{Geometry and Density Operators}

In 1985, Campbell showed that geometry can be introduced into probability
calculus as follows \cite{campbell}: for a fixed probability distribution,
define the inner product of two random variables to be the expectation of
the product of these variables. Differential geometry emerges when we
consider varying the probability distribution, either directly or through
changing parameters on which the distribution depends. Within such a
geometric framework, the sets of probability distributions are viewed as
differentiable manifolds, the random variables appear as vectors and the
expectation values of random variables are replaced with inner products in
tangent spaces to such manifolds of probabilities. In particular, the MaxEnt
estimate of a probability distribution given a prior and the information
constraints is found by following an integral curve through the prior which
is orthogonal (in the Fisher metric on the simplex) to the hyperplane
defined by the information constraint equation until the constraint is
satisfied. In 1995, Braunstein and Caves extended Campbell's ideas to the
quantum framework \cite{sam}. Here, following their analysis, we provide a
differential geometric viewpoint of the (generalized) quantum MaxEnt
estimate (\ref{maxent-estimate}).

Consider the quantum analogue $\mathcal{M}_{\vec{\rho}}$ of the probability
simplex, the space of density operators $\vec{\rho}$ written as vectors in $%
\mathcal{L}\left( \mathcal{H}\right) $, the linear space of all linear
operators on a $n$-dimensional Hilbert space $\mathcal{H}$,%
\begin{equation}
\mathcal{M}_{\vec{\rho}}\overset{\text{def}}{=}\left\{ \vec{\rho}\in 
\mathcal{L}\left( \mathcal{H}\right) :\vec{\rho}\overset{\text{def}}{=}%
\sum_{i\text{, }j=1}^{n}\rho ^{ij}\vec{e}_{ij}\text{, }\vec{\rho}=\vec{\rho}%
^{\dagger }\text{, tr}\left( \vec{\rho}\right) =1\text{, }\vec{\rho}\text{ }%
\geq 0\right\} \text{.}
\end{equation}%
The space $\mathcal{M}_{\vec{\rho}}$ is an $\left( n^{2}-1\right) $%
-dimensional \emph{real} manifold with complicated boundary. An arbitrary
linear operator vector $\vec{V}$ on $\mathcal{H}$ can be decomposed in terms
of an operator vector basis $\vec{e}_{ij}\overset{\text{def}}{=}\left\vert
i\right\rangle \left\langle j\right\vert $ with $i$, $j=1$,..., $n$ as
follows,%
\begin{equation}
\vec{V}=\sum_{i\text{, }j=1}^{n}\left\langle i|\vec{V}|j\right\rangle \vec{e}%
_{ij}\text{ }=\sum_{i\text{, }j=1}^{n}V^{ij}\vec{e}_{ij}\text{ .}
\end{equation}%
The tangent space at $\vec{\rho}$ is an $\left( n^{2}-1\right) $-dimensional 
\emph{real} vector space of traceless Hermitian operators $\vec{T}$,%
\begin{equation}
\vec{T}=\sum_{i\text{, }j=1}^{n}T^{ij}\vec{e}_{ij}\text{, tr}\left( \vec{T}%
\right) =0\text{.}
\end{equation}%
The action of $1$-forms $\tilde{F}$ expanded in terms of the dual basis $%
\tilde{\omega}^{ji}\overset{\text{def}}{=}\left\vert i\right\rangle
\left\langle j\right\vert $,%
\begin{equation}
\tilde{F}\overset{\text{def}}{=}\sum_{i\text{, }j=1}^{n}F_{ij}\tilde{\omega}%
^{ji}\text{,}
\end{equation}%
on density operators $\vec{\rho}$ is defined as follows,%
\begin{equation}
\tilde{F}\left( \vec{\rho}\right) \equiv \left\langle \tilde{F}\text{, }\vec{%
\rho}\right\rangle =\sum_{i\text{, }j\text{, }l\text{, }k=1}^{n}F_{ij}\rho
^{lk}\left\langle \tilde{\omega}^{ji}\text{, }\vec{e}_{lk}\right\rangle
=\sum_{i\text{, }j\text{, }l\text{, }k=1}^{n}F_{ij}\rho ^{lk}\delta
_{l}^{j}\delta _{k}^{i}=\sum_{i\text{, }j=1}^{n}F_{ij}\rho ^{ji}=\text{tr}%
\left( \tilde{F}\vec{\rho}\right) \equiv \left\langle \tilde{F}\right\rangle 
\text{.}
\end{equation}%
Therefore, an Hermitian $1$-form $\tilde{F}=\tilde{F}^{\dagger }$ is an
ordinary quantum observable with $\left\langle \tilde{F}\text{, }\vec{\rho}%
\right\rangle =\left\langle \tilde{F}\right\rangle $. A metric structure $g_{%
\vec{\rho}}\left( \cdot \text{, }\cdot \right) $ on the manifold $\mathcal{M}%
_{\vec{\rho}}$ can be introduced by defining the metric's action on a pair
of $1$-forms $\tilde{A}$ and $\tilde{B}$ as follows,%
\begin{equation}
g_{\vec{\rho}}\left( \tilde{A}\text{, }\tilde{B}\right) \overset{\text{def}}{%
=}\left\langle \frac{\tilde{A}\tilde{B}+\tilde{B}\tilde{A}}{2}\right\rangle =%
\text{tr}\left[ \left( \frac{\tilde{A}\tilde{B}+\tilde{B}\tilde{A}}{2}%
\right) \vec{\rho}\right] =\text{tr}\left[ \frac{\tilde{A}}{2}\left( \vec{%
\rho}\tilde{B}+\tilde{B}\vec{\rho}\right) \right] =\left\langle \tilde{A}%
\text{, }\mathcal{R}_{\vec{\rho}}\left( \tilde{B}\right) \right\rangle \text{%
,}
\end{equation}%
where $\mathcal{R}_{\vec{\rho}}\left( \tilde{B}\right) $ is the raising
operator mapping $1$-forms (lower covariant components) to vectors (upper
contravariant components), 
\begin{equation}
\mathcal{R}_{\vec{\rho}}\left( \tilde{B}\right) \overset{\text{def}}{=}\frac{%
\vec{\rho}\tilde{B}+\tilde{B}\vec{\rho}}{2}\text{.}
\end{equation}%
Such a metric is formulated in terms of statistical correlations of quantum
observables. Using the lowering operator $\mathcal{L}_{\vec{\rho}}\left( 
\vec{A}\right) $ mapping vectors to $1$-forms,%
\begin{equation}
\mathcal{L}_{\vec{\rho}}\left( \vec{A}\right) =\mathcal{R}_{\vec{\rho}%
}^{-1}\left( \vec{A}\right) \text{,}
\end{equation}%
we can also define the action of the metric tensor $g_{\vec{\rho}}\left(
\cdot \text{, }\cdot \right) $ on a pair of vectors $\vec{A}$ and $\vec{B}$,%
\begin{equation}
g_{\vec{\rho}}\left( \vec{A}\text{, }\vec{B}\right) \overset{\text{def}}{=}%
\left\langle \mathcal{L}_{\vec{\rho}}\left( \vec{A}\right) \text{, }\vec{B}%
\right\rangle =\text{tr}\left[ \vec{B}\mathcal{L}_{\vec{\rho}}\left( \vec{A}%
\right) \right] \text{.}
\end{equation}%
The quantum line element $ds^{2}$,%
\begin{equation}
ds^{2}=g_{\vec{\rho}}\left( d\vec{\rho}\text{, }d\vec{\rho}\right) \text{,}
\end{equation}%
with $d\vec{\rho}$ given by,%
\begin{equation}
d\vec{\rho}=\sum_{j=1}^{n}dp^{j}\left\vert j\right\rangle \left\langle
j\right\vert +id\theta \sum_{m\text{,}\ l=1}^{n}\left( p^{m}-p^{l}\right)
h_{lm}\left\vert l\right\rangle \left\langle m\right\vert \text{,}
\end{equation}%
and with $e^{id\theta h}$ an infinitesimal unitary transformation on the
orthonormal basis that diagonalizes $\vec{\rho}$, reads%
\begin{equation}
ds^{2}=\sum_{k=1}^{n}\frac{\left( dp^{k}\right) ^{2}}{p^{k}}+2d\theta
^{2}\sum_{j\neq k}\frac{\left( p^{j}-p^{k}\right) ^{2}}{\left(
p^{j}+p^{k}\right) }\left\vert h_{jk}\right\vert ^{2}\text{.}  \label{qle}
\end{equation}%
Notice that the quantum line element (\ref{qle}) is identical to the
distinguishability metric for density operators obtained by Braunstein and
Caves by optimizing over all generalized quantum measurements for
distinguishing among neighboring quantum states \cite{samPRL}.

The notion of quantum "transport" can also be introduced in this geometric
framework. Consider a surface defined by the relation $\left\langle \Delta 
\tilde{A}\right\rangle =0$, where $\Delta \tilde{A}\overset{\text{def}}{=}%
\tilde{A}-\left\langle \tilde{A}\text{, }\vec{\rho}\right\rangle \tilde{1}$
is the zero mean observable and $\tilde{1}$ is the unit $1$-form operator.
Any tangent vector $\vec{t}$ that lies on the surface $\left\langle \Delta 
\tilde{A}\right\rangle =0$ satisfies $\left\langle \Delta \tilde{A}\text{, }%
\vec{t}\right\rangle =0$. This implies that a vector field $\mathcal{R}_{%
\vec{\rho}}\left( \Delta \tilde{A}\right) $ associated with the observable $%
\Delta \tilde{A}$ is orthogonal to the surface $\left\langle \Delta \tilde{A}%
\right\rangle =0$ since,%
\begin{equation}
g_{\vec{\rho}}\left( \mathcal{R}_{\vec{\rho}}\left( \Delta \tilde{A}\right) 
\text{, }\vec{t}\right) =\left\langle \Delta \tilde{A}\text{, }\vec{t}%
\right\rangle =0\text{.}
\end{equation}%
The most efficient way to construct a path that goes from the state $\vec{%
\rho}_{0}$ to another state that has a different expectation value for $%
\tilde{A}$ is by consistently moving in the direction orthogonal to the
surfaces $\left\langle \Delta \tilde{A}\right\rangle =0$. This navigation is
accomplished by selecting a parametrized path $\vec{\rho}=\vec{\rho}\left(
\lambda \right) $ with $\vec{\rho}\left( \lambda =0\right) =\vec{\rho}_{0}$
along the vector field $\mathcal{R}_{\vec{\rho}}\left( \Delta \tilde{A}%
\right) $ such that,%
\begin{equation}
\frac{d\vec{\rho}\left( \lambda \right) }{d\lambda }+\mathcal{R}_{\vec{\rho}%
}\left( \tilde{A}-\left\langle \tilde{A}\right\rangle \tilde{1}\right) =0%
\text{,}
\end{equation}%
that is,%
\begin{equation}
\vec{\rho}\left( \lambda \right) =\frac{e^{-\frac{\lambda }{2}\tilde{A}}\vec{%
\rho}_{0}e^{-\frac{\lambda }{2}\tilde{A}}}{\text{tr}\left( e^{-\lambda 
\tilde{A}}\vec{\rho}_{0}\right) }\text{.}  \label{quantum estimate}
\end{equation}%
The path (\ref{quantum estimate}) is reminiscent of a quantum exponential
model of \emph{symmetric} type \cite{oe}. In conclusion, from a differential
geometric point of view, the (generalized) MaxEnt estimate (\ref%
{maxent-estimate}) can be regarded as a quantum trajectory (\ref{quantum
estimate}) that passes orthogonally through the family of surfaces of
constant expectation value corresponding to some observable ($1$-form).

\section{Final Remarks}

In this article, we presented a differential geometric viewpoint of the
quantum MaxEnt estimate of a density operator when only incomplete knowledge
encoded in the expectation values of a set of quantum observables is
available. The additional possibility of considering some prior bias towards
a certain density operator (the prior) was taken into account by using MrE
and the unsolved issues with its relative entropic inference criterion were
pointed out.

Although we limited our analysis to uniform priors and commutative
observables (non-commutative extensions of (\ref{maxent-estimate}) are under
investigation), we must point out that care is needed in handling the
non-commutativity of quantum observables. Furthermore, Jaynes' quantum
MaxEnt method deals solely with \emph{assigning} density operators. In the
most optimistic scenario, we could use this method also in the presence of
uniform priors. However, in actual scenarios, the prior is not uniform and a
quantum entropic \emph{updating} method is required. Unfortunately, there is
a lack of a direct axiomatic justification (consistency requirements) for
the use of the logarithmic quantum relative entropy maximization criterion
for entropic inferences \cite{bala}. We also stress that although the
Maximum relative Entropy method for updating probabilities includes both
Jaynes' MaxEnt and Bayes' rule \cite{caticha-giffin}, a similar result in
the quantum framework is still missing. Finally, Bayesian quantum inferences
for determining quantum states depend on the prior as well as measurement
data, even in the limit of an infinite number of measurements \cite{fuchs}.
Furthermore, we are also aware that there has been some controversy over the
naturalness of relative entropy as a tool for quantum statistical inference 
\cite{donald}. As a matter of fact, when the quantum-mechanical concept of
relative entropy is discussed from an information-theoretic point of view,
it can be shown that not all definitions found in the literature are equally
suitable for the purpose of statistical inference by entropy maximization 
\cite{beno}.

In view of all the above-mentioned considerations, we believe that an
axiomatic formulation of a "universal" quantum entropic updating methodology
embracing the quantum Bayes' rule \cite{schack} as a special case should to
be developed. However, just as the Maximum relative Entropy method did this
for the classical case, it is our hope that a quantum version (with
non-uniform priors) can be constructed.

\begin{acknowledgments}
C. Cafaro thanks Ariel Caticha for very helpful discussions. The research of
C. Cafaro, C. Lupo and S. Mancini has received funding from the European
Commission's Seventh Framework Programme (FP7/2007--2013) under grant
agreements no. 213681.
\end{acknowledgments}


\begin{thebibliography}{99}
\bibitem{frankel} T. Frankel, "\emph{The Geometry of Physics}", Cambridge
University Press (1997).

\bibitem{jaynes1} E. T. Jaynes, "\emph{Macroscopic Prediction}", in Complex
Systems and Operational Approaches in Neurobiology, Physics, and Computers,
ed. by H. Haken, Springer-Berlin, (1985).

\bibitem{brody} D. C. Brody and L. P. Hughston, Phys. Rev. Lett. \textbf{77}%
, 2851 (1996).

\bibitem{skilling} J. Skilling, AIP Conf. Proc. \textbf{954}, 39 (2007).

\bibitem{amari} S. Amari and H. Nagaoka, "\emph{Methods of Information
Geometry}", Oxford University Press (2000).

\bibitem{caticha-giffin} A. Caticha and A. Giffin, AIP Conf. Proc. \textbf{%
872}, 31 (2006); A. Giffin and A. Caticha, AIP Conf. Proc. \textbf{954}, 74
(2007).

\bibitem{jay} E. T. Jaynes, Phys. Rev. \textbf{106}, 620 (1957).

\bibitem{carlo} C. Cafaro, \textit{Chaos, Solitons \& Fractals} \textbf{41},
886 (2009).

\bibitem{jaynes2} E. T. Jaynes, Phys. Rev. \textbf{108}, 171 (1957).

\bibitem{grasselli} M. R. Grasselli, "\emph{Classical and Quantum
Information Geometry}", PhD Thesis, King's College London (2001).

\bibitem{bala} R. Balian and N. L. Balazs, Annals of Physics \textbf{179},
97 (1987).

\bibitem{private} A. Caticha, "\emph{Notes on Quantum Relative Entropy}",
unpublished, University at Albany - SUNY (2004).

\bibitem{buzek} V. Buzek, G. Drobny, R. Derka, G. Adam and H. Wiedemann, 
\textit{Chaos, Solitons \& Fractals} \textbf{10}, 981 (1999).

\bibitem{schack} R. Schack, T. A. Brun and C. M. Caves, Phys. Rev. \textbf{%
A64}, 014305 (2001).

\bibitem{giffin} A. Giffin, "\emph{Maximum Entropy: The Universal Method for
Inference}", Ph. D. Thesis, SUNY at Albany, Albany (New York- USA),
http://arxiv.org/abs/0901.2987 (2008).

\bibitem{campbell} L. L. Campbell, Inform. Sci. \textbf{35}, 199 (1985).

\bibitem{sam} S. L. Braunstein and C. M. Caves, Annals of the New York
Academy of Sciences \textbf{755, }786 (1995).

\bibitem{samPRL} S. L. Braunstein and C. M. Caves, Phys. Rev. Lett. \textbf{%
72}, 3439 (1994).

\bibitem{oe} O. E. Barndorff-Nielsen, R. D. Gill and P. E. Jupp, J. Roy.
Soc. \textbf{B65}, 775 (2003).

\bibitem{fuchs} C. A. Fuchs and R. Schack, AIP Conf. Proc. \textbf{1101},
255 (2009).

\bibitem{donald} M. J. Donald, Commun. Math. Phys. \textbf{105}, 13 (1968).

\bibitem{beno} R. W. Benoist, J.-P. Marchand and W. Wyss, Lett. Math. Phys. 
\textbf{3}, 169 (1979).
\end{thebibliography}
\end{document}